\documentclass[conference]{IEEEtran}
\IEEEoverridecommandlockouts
\usepackage{cite}
\usepackage{amsmath,amssymb,amsfonts}
\usepackage{algorithmic}
\usepackage{graphicx}
\usepackage{textcomp}
\usepackage{xcolor}
\usepackage{adjustbox}
\def\BibTeX{{\rm B\kern-.05em{\sc i\kern-.025em b}\kern-.08em
    T\kern-.1667em\lower.7ex\hbox{E}\kern-.125emX}}

\graphicspath{{figures/}}
\begin{document}

\title{On Batching Acknowledgements in C-V2X Services\\
\thanks{This work is in part supported by Ford Motor Company}
}

% AUTHOR LIST
\author{\IEEEauthorblockN{Mahdi Zaman, Md Saifuddin, Mahdi Razzaghpour}
\IEEEauthorblockA{\textit{University of Central Florida} \\
% \textit{name of organization (of Aff.)}\\
% City, Country \\
\{mahdizaman, md.saif, razzaghpour.mahdi\}@knights.ucf.edu}
\and
% \IEEEauthorblockN{Md Saifuddin}
% \IEEEauthorblockA{\textit{University of Central Florida} \\
% % \textit{name of organization (of Aff.)}\\
% % City, Country \\
% md.saif@knights.ucf.edu}
\and
% \IEEEauthorblockN{Ghayoor Shah}
% \IEEEauthorblockA{\textit{University of Central Florida} \\
% % \textit{name of organization (of Aff.)}\\
% % City, Country \\
% gshah8@knights.ucf.edu}
% \and
\IEEEauthorblockN{Yaser Fallah}
\IEEEauthorblockA{\textit{University of Central Florida} \\
% \textit{name of organization (of Aff.)}\\
% City, Country \\
yaser.fallah@ucf.edu}
\and
\IEEEauthorblockN{Jayanthi Rao}
\IEEEauthorblockA{\textit{Ford Motor Company} \\
% \textit{name of organization (of Aff.)}\\
% City, Country \\
jrao1@ford.com}
% \and
% \IEEEauthorblockN{5\textsuperscript{th} Given Name Surname}
% \IEEEauthorblockA{\textit{dept. name of organization (of Aff.)} \\
% \textit{name of organization (of Aff.)}\\
% City, Country \\
% email address or ORCID}
% \and
% \IEEEauthorblockN{6\textsuperscript{th} Given Name Surname}
% \IEEEauthorblockA{\textit{dept. name of organization (of Aff.)} \\
% \textit{name of organization (of Aff.)}\\
% City, Country \\
% email address or ORCID}
}

\maketitle

\begin{abstract}
Cellular Vehicle-to-Everything (C-V2X) is a frontier in the evolution of distributed communication introduced in 3GPP release 14 to advanced use cases. While research efforts continue to optimize the accessible bandwidth for transportation ecosystem, a bottom up analysis from the application layer perspective is necessary prior to deployment, as it can expose potential issues that can emerge in a dynamic road environment. This emphasizes on assessing the network using application-oriented metrics to evaluate its capacity of providing advanced vehicular services with stringent latency and throughput requirements. C-V2X enables advanced applications like autonomous driving and on-the-go transaction services where consecutive exchange of messages is required. For such services, the network level metrics fails to capture the edge case service quality as they express an average measure of performance. In this paper, we present an application-oriented analysis of a transaction service built on C-V2X protocol. We analyze different design choices that affects quality of service both from network-oriented and user-centric metrics and we highlight the issues regarding packet dissemination from infrastructures for vehicle-to-infrastructure (V2I) based service applications. We also present our study on  the impact of batching in disseminating acknowledgement packets (ACK) and its consequence on both the service reliability and network congestion. Our results show that time-sensitive and mission-sensitive vehicular applications should aim for a balance between achieving the mission utility in shortest duration possible, while keeping minimal impact on the system-wide stability.
\end{abstract}

\section{Introduction}
Vehicle-to-everything (V2X) communication is expected to connect the road-user entities in the larger Internet-of-Things (IoT) network to ensure safe and seamless service for the users. Research and testing activities have shown the efficiency of Cellular V2X (C-V2X) in maintaining high reliability \& low latency communication in vehicular environment. Any entity equipped with C-V2X User Equipment (UE) can utilize connectivity via C-V2X protocol. This enables vehicular user units to communicate with other vehicles (V2V), pedestrians (V2P), networks (V2N) and infrastructures (V2I). The early design choices for V2X-based services primarily aimed to ensure robust and efficient communication between mobile entities. C-V2X, with the underlying support of Long Term Evolution (LTE) constructs the core of such communication to facilitate its participants with the capability of broadcasting a snapshot of their dynamic state information to the neighboring circle. Alongside, its perceived robustness and scalability to host large number of moving users with or without infrastructural support makes connectivity an indisputable component for the envisioned autonomy in transportation.  

While under network coverage, C-V2X conducts physical layer assessment and resource allocation via centralized authority, ie the eNodeB accesses information about the participants within its cell and allocates resources. In contrast, C-V2X generally assumes the same functionalities under no network coverage. In this mode, resource allocation is performed independently by the entities in a distributed fashion with limited information based on individual receptions. This distributed mode of operation, also known as mode-4, was initiated in 3GPP release 14  which C-V2X utilizes for broadcasting Basic Safety Messages (BSM). Periodic transmission and reception of BSMs functionally allow vehicles to perceive longer range of surroundings and thus enhance safety at high mobility. However, the scalability analyses on C-V2X \cite{toghi2019analysis} as well as the additional resource allowance \cite{fccNov2020} strengthens C-V2X to accommodate advanced safety applications namely cooperative platooning, teleoperated driving etc. This does not only pave the way for autonomous driving, but also provides a way to remodel the traditional traffic services with more efficient and cost-effective deployment. For research objectives, this brings a change to the way network performance is trivially analyzed.

Safety services can rely on periodic BSM transmissions often aided by a congestion control method \cite{toghi2019analysis} at heavy traffic. BSM consists state information about a vehicle. But it is often more appropriate to conduct the advanced applications through context-aware messages. The set of advanced use cases laid out by 5GAA \cite{5gaa_use_cases_2020} such as awareness confirmation, vehicle decision assist, coordinated cooperative driving manoeuvre \cite{Razz2204:Finite}, HD map sharing\cite{electronics11203374} and so forth describes scenarios where context-specific messages can achieve the mission with higher accuracy, replacing the need for BSM transmission for such purposes. While interoperability between these services require standardization, the specific application and method design are still under research and development. From the resource allocation perspective, the application-specific messages can differ from BSM in several ways. Firstly, unlike BSMs, these messages can be aperiodic. Secondly, the mission of these messages are to initiate and confirm a particular usage, hence multiple message exchange can be required between two specific entities (a host and remote vehicle pair, or an RSU and a remote vehicle). Thirdly, these messages are expected to be assigned a lower priority so that the reception of BSMs are never affected. % lead this para to basic-to-advanced safety requirement changes, aperiodic-periodic, multiple priority, multiple transmission etc. then specify what specific contributions are made in this paper. 

%For ensuring Quality of Service (QoS) in these communications, research effort has long been invested to tackle the unique challenges that can appear in a vehicular scenario; primarily high-speed mobility and interference. 3GPP release 14 has prioritized the basic safety use cases and utilizes periodic Basic Safety Messages (BSM) for such services. Thus release 14 primarily lays out the resource selection abstraction and the application layer functions for periodic communication. On the other hand, for the advanced set of applications namely cooperative platooning, teleoperated driving, dynamic charging etc, the communication occurs through consecutive message exchanges which are aperiodic in nature. In addition, BSM based use cases are subject to a QoS requirement of 100ms latency and 80-90\% reliability, while advanced use cases that enable higher degree of automation has more stringent requirements as 10ms latency and 99.99\% reliability \cite{3gpp36885_study_v2x_services}. 
%This calls for specific analyses on service performances where aperiodic packets are utilized. 

To ensure the stringent QoS requirement \cite{5gaa_use_cases_2020} for the advanced application services under distributed mode, aperiodic varying priority messages need to efficiently utilize the C-V2X resource management architecture. Using a reference scenario introduced in \cite{mzaman_trigger}, we study a vehicular service where multiple messages are exchanged in sequence to complete a service utility. Under similar context of communication, the reference can be stretched for any two C-V2X equipped entities where one entity acts as the service manager. Our contributions in this paper is threefold: we identify batchsize as a crucial design factor in both the  vehicular service performance as well as on the resource consumption, we show that application services require usage of adaptive batchsize to maintain scalable and reliable service performance keeping safety as the prime objective, and we present the performance analysis results in terms of application-centric metrics to explain the limitations of C-V2X architecture in providing aperiodic communication that captures the edge case performance as well as the holistic measure.

\section{Related Works}
Utilizing communication systems for efficient traffic management has been of interest from researchers since long. Early works by Schulz et al. \cite{schulz1996traffic} discusses the features, benefits and drawbacks of the concurrent communication technologies for this purpose. This work sheds light on the radio data system, GSM, Short Messaging Service (SMS), general packet radio service (GPRS) for traffic management. Although these protocols proved the potential for providing autonomous in-car navigation, it was not sufficiently scalable for high density traffic. Among early works, Djahel et al \cite{djahel2013adaptive} discusses a protocol-agnostic adaptive traffic management architecture for emergency vehicles with ability to adjust per driving policies, driver behavioral change and essential security control. 

As communication protocols evolved, DSRC and later LTE grounded the longstanding expectations of achieving a scalable and reliable vehicular communication. Using the dedicated ITS bandwidth of 75MHz, Dedicated Short Range Communication (DSRC) first enabled direct vehicle-to-vehicle (V2V) communication via distributed mode of operation \cite{Kenney_DSRC}. While DSRC primarily aims to provide low latency in safety applications, LTE-V2X emerged with a promise of longer range and higher interoperability \cite{Naik_dsrc_cv2x_comparison}. Subsequent standard releases by 3GPP enabled C-V2X to be the major contender of the spectrum while the design focus remained on ensuring low latency and high reliability in safety applications. In addition, 3GPP \cite{3gpp36885_study_v2x_services} laid out advanced sets of use cases that would enable the perceptual needs for connected automated vehicles via sensor sharing; network-assisted teleoperated driving, V2P-enabled pedestrian safety, V2I-assisted toll collection to name a few. 
Besides the advancement on mobility support, the envisioned applications have been explored by some works that shows the feasibility of the concurrent state of C-V2X Vehicular User Equipments (VUE) in terms of deployment. 

In advanced V2X applications like cooperative driving and sensor-data sharing, entities exchange messages (V2V or V2I) for arbitration, whereas fee-collection services involve message exchange (V2I) for transaction; both type requiring consecutive message exchange between two entities. Since each sequence of message have one specific mission to achieve, these are generally aperiodic packets where each message in the sequence In \cite{lusvarghi2020coexistence}, the authors show the limitations of the current C-V2X protocol in handling aperiodic packets, specially when the traffic arrival rate is high or the aperiodic packets are large, which can be the case to meet the data needs for advanced applications. Hence the advanced use cases require defining new message set dictionary \cite{saej3224,saej3217}. For instance,  Fine-tuning of the channel parameters may be required to accommodate both periodic and aperiodic packets in the same channel.

It is evident that C-V2X needs a application-oriented evaluation to address the afresh concerns. Authors in \cite{wendland2019application} suggests amendments in the mode-4 application layer for resource selection scheme and emphasizes on the need for application-oriented evaluation to maintain efficient cooperative awareness. The authors provides insight on the limited performance of Semi Persistent Scheduling (SPS) mechanism when resources are reused in concurrent iteration of selection. In presence of aperiodic packets in the channel, it can threaten the SPS efficiency due to similar event as reuse of resources. To utilize C-V2X links for service applications, it is important to address these potential threats for robust and safe driving maneuver.

In previous work \cite{mzaman_trigger}, we present an infrastructure-based service protocol that provides generalized transaction services over C-V2X RAN. We experiment with different design parameters of the protocol to present an optimized set of procedures that ensures both safety and quality of service for the users. In this work, we elaborate the analysis of the protocol keeping the focus on the scalability issues for such services; such as acknowledge packet (ACK) management from the infrastructure entity. We experiment with different batching policies and discuss the potential impact under different traffic flow. As a whole, the goal is to design the transaction protocol towards a deployable format with deterministic decisions on the parameters. To the best of authors' knowledge, no past analyses presents the scalability issues on the advanced service set where C-V2X accommodates both periodic and aperiodic packets of varying priority on  the latest standardized architecture. As different services offer different utility and thus different QoS requirement, a predictive QoS approach \cite{boban2021predictive} can benefit interoperability between these services. In light of the service structure presented in this paper, we hypothesize that an adaptive scheme to adjust queue length as a function of concurrent traffic flow may ensure reduced wait time for service users as well as reduced resource consumption in realistic driving conditions, thus offering higher predictability in the quality of service.

\begin{figure}[b]
\centerline{\includegraphics[width=0.52\textwidth,height=5.5cm]{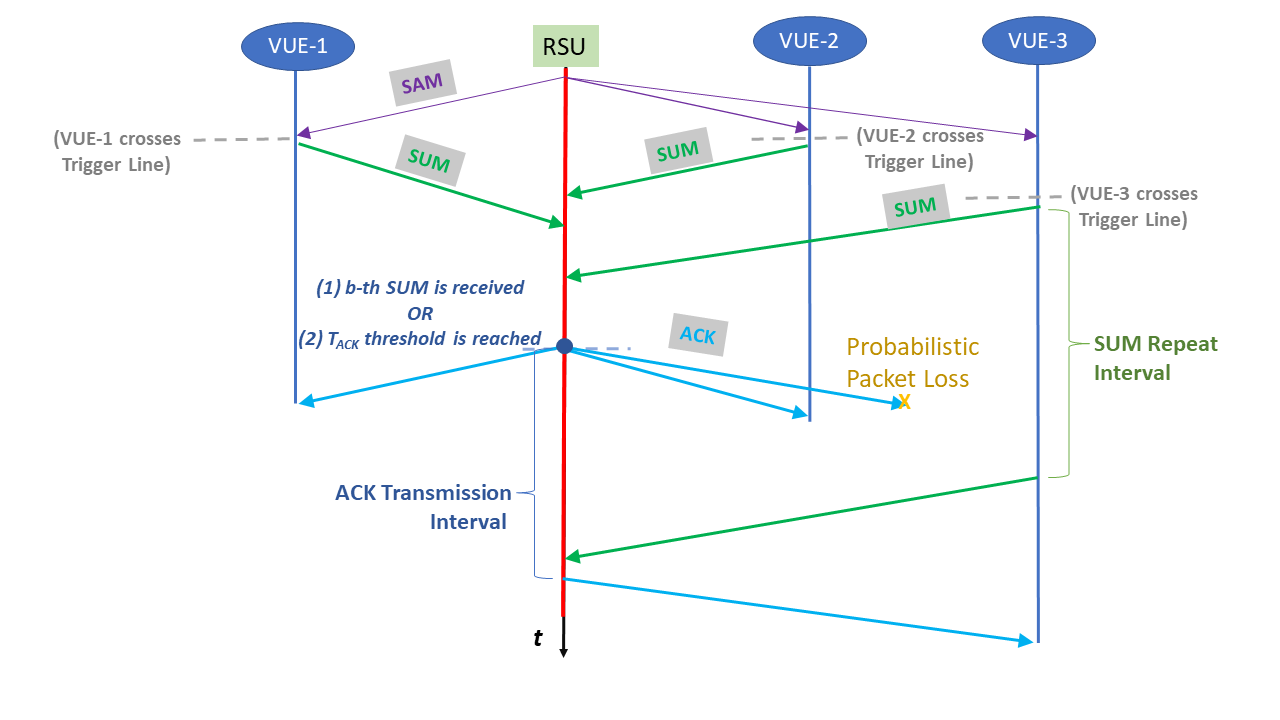}}
% \centerline[width=0.5\textwidth,height=6cm]{cbr-15_20_30.png}
\caption{Service Procedure Timeline}
\label{fig_service_procedure}
\end{figure}

\section{System Model}
As a reference application, we utilize a V2I-based transaction service. Figure \ref{fig_service_procedure} demonstrates the exchange of service messages on a timeline; as outlined in the following portion. We assume a freeway where an RSU acts as the service provider, which broadcasts a $700 byte$ Service Advertisement Packet (SAM). SAM consists necessary information for subscribed users to respond to based on their eligibility. A subscribed VUE can respond to the advertisement by broadcasting a usage request (SUM) to the RSU when they cross a virtual trigger line. Our previous work shows the optimal trigger distance to be 0m \cite{mzaman_trigger}, which is the value we have used for the simulation results presented in this paper. Upon checking the VUE's eligibility of usage, subcription status, user authentication etc, the RSU then broadcasts a response packet (ACK) with necessary information to declare the usage as complete for the specific set of vehicles. Upon reception of ACK, a VUE considers the service as completed, and ends the relevant procedures. Until a VUE receives an ACK, it continues to transmit SUM periodically every $600ms$. Depending on ACK batchsize ($b$), one ACK can be broadcast to address a single VUE or a group of VUEs that has interacted within a preceding duration. On the RSU side, all SUM receptions are treated with equal opportunity, making packet loss the primary source of randomness in reception. However, as mentioned before, the V2I packets carry a lower priority tag than BSMs \cite{saej3217} which affects the transmission queue if a BSM and a SUM is concurrently in it. This can occur only on the VUE's side (because BSM concerns safety broadcasts and hence is not a concern for a stationary RSU), and the V2I packet in the concurrent queue is transmitted in the next available transmission opportunity so we assume the impact of this event to be minimal. To further strengthen the reception probabilities, V2I packets adopt hybrid automatic repeat request (HARQ) similar to BSM transmissions, which enables a packet scheduled at $t$ to be retransmitted at a randomly chosen subframe within [$t-15$, $t+15$] window. 

At each step of these message exchanges, a series of authentication subroutines (management, verification and approval) are involved. However, in this article, we only focus on the network-level analyses, and so these subroutines are estimated as time headroom between successive packet exchanges alongside the intra-layer transmission delay. It is important to mention that unlike mode-3 where an RSU can provide functionalities as an eNodeB, our tests assume mode-4 communication with distributed resource selection mechanism. Hence the prototype RSU is only responsible for management of the service. It is assumed that the RSU carries out the front-end functions of specific services following 3GPP specifications \cite{3gpp23700_APPenhancementV2X}, whereas the back-end computations for secure handling of private data can be executed on-device or offloaded to edge nodes.

\section{Experiment Setup}

\begin{table}[t] 
% \hspace*{-4cm}\centering
\caption{Simulation Parameters \& Configurations}
\begin{center}
\bgroup
\def\arraystretch{1.1}
\begin{tabular*}{0.46\textwidth}{@{\extracolsep{\fill} }  l r }
\hline
\hline
Roadlength  & 3.0 km\\
No. of lanes & 16 \\
Traffic Crossing Rate & 10, 15, 20 \& 30 veh/sec \\
Propagation Loss Model & I-405 Model \cite{ehsanChmodel} \\
\hline
Simulation Time    & 50 second \\ 
Carrier Freq.   & 5.905 GHz\\     
Bandwidth & 20.00 MHz\\ 
BSM Transmission Periodicity & [100-600] ms \\
CBR Threshold & -92 dBm \\
Congestion control & Enabled for BSM \\
One-shot Transmission & Enabled for BSM \\
Hybrid Automatic Repeat Request & Enabled for all packets \\
\hline
SAM Payload Size   & 700 Bytes \\ 
SAM MCS & 7 \\ 
SUM Payload Size   & 450 Bytes \\ 
SUM MCS & 11 \\ 
ACK Payload Size   & 300 Bytes \\
ACK MCS & 6 \\ 
SAM periodicity & 1s \\
Trigger Distance & 0m \\
SUM Repeat Interval & 600ms \\
ACK Transmission Interval & 400ms \\
BSM ProSe Per Packet Priority & 2 \\
SAM/SUM/ACK ProSe Per Packet Priority & 6 \\
\hline
\end{tabular*}
\egroup
\label{table:tollingSpecs}
\end{center}
\end{table}

\subsection{Scenario Description}
We deployed the service prototype in a link-level network simulator equipped with C-V2X protocol layers. The hi-fidelity system level simulator has been developed over several years and validated with real-world data with specific focus on scalability \cite{Toghi_multiple}. The base framework to emulate Device-to-device communication is founded on ns3 equipped with D2D modules \cite{rouil2017implementationD2D} and enhanced to fit the modifications that allow C-V2X system simulation as per 3GPP rel 14. 

We designed a 3km, 16 lanes bidirectional freeway with an RSU situated in the middle of the stretch. The RSU functions as the service provider for traffic along both directions. For testing the protocol performance under different traffic conditions, we distribute varying number of moving vehicles uniformly over the 3km road. All vehicles are equipped with C-V2X vehicular user equipment (VUE), who are capable of exchanging BSM (among themselves) and service packets (with the RSU). For the purpose of this study, we characterize traffic density in terms of traffic flow rate across the trigger line in units of vehicle per second. Since every eligible VUE transmits a SUM at the time of crossing trigger line, this quantity tracks the rate of base SUM transmission every second. However, the total number of SUM transmission per second can be higher due to the repeated SUM attempts, which is elaborated in the following subsection. \\

% % \begin{table}
% % \caption{Traffic Flow Rate}
% \begin{center}
% % \begin{tabular}{ |P{3cm}||P{3cm}|  }
% \begin{tabular}{ |P{3cm}|P{3cm}  }
%  \hline
%   \multicolumn{2}{|c|}{Traffic Flow Rate} \\
%  \hline
%  Vehicle Per Second & Category\\
%  \hline
% 1   & very low \\
% 5   & low \\
% 10   & low-medium \\
% 15   & medium \\
% 20   & high \\
% 30   & very high \\
%  \hline
% \end{tabular}
% \end{center}
% % \end{table}

\begin{center}
\begin{tabular}{|cc|}
\hline \multicolumn{2}{|c|}{ Traffic Flow Rate } \\
\hline Vehicle Per Second & Category \\
\hline 1 & very low \\
5 & low \\
10 & low-medium \\
15 & medium \\
20 & high \\
30 & very high \\
\hline
\end{tabular}
\end{center}

\subsection{Performance Metrics}
\subsubsection{Service Completion Time}
The key performance metric for the service performance is Service Completion Time (SCT). It is defined as the interval between a vehicle's SUM transmission, and an ACK reception from the RSU. SCT is measured from the VUE end to ensure a user-centric assessment. Each vehicle and the RSU in our scenario logs its own timestamps relevant for the V2I packet transmission and reception. Hence this incorporates the scheduling delay on both transmission ends. In C-V2X the scheduling delay is the delay induced by Medium Access Control (MAC) layer when a packet generation request has arrived from the service layer and a suitable resource is being located via SPS procedure at the MAC layer. This resource allocation procedure takes on the range of [4 100] ms on each end.  Since SCT measures the end-to-end duration of at least two packets, two unit of propagation delay adds up to the net total SCT. Hence for one complete SCT computation cycle starting with transmitting a request and ending with an ACK reception, a VUE can be subject to at least the sum of the two scheduling delays, i.e. in the range of [8 200] ms. Considering HARQ-supported worst case reception, this sum can reach upto [$8$ $230$] ms.
\subsubsection{Attempt Count}
We tracked the number of SUM attempts required by individual VUEs under different scenarios and refer to it as Attempt Count (AC). We acknowledge that AC gauges SCT in a quantized fashion where, due to the protocol design, each additional attempt accrues 600ms delay in SCT. Hence reducing the number of attempts required to complete the transaction is desirable. Although it is difficult to selectively facilitate V2I receptions to strengthen service performance, specific attention on AC can be worthwhile, as maximizing the reception at early attempts can reduce following cascade of losses. 
\subsubsection{Packet Error Rate}
We observed Packet Error Rate (PER) for BSMs to measure the impact of the V2I service on the overall network performance so that safety is retained. PER is measured in trivial ways as the ratio of percentage of successfully received packets and the percentage of all transmitted packets within a certain vicinity. It provides an average measure of the channel condition to identify potential causes and impact. The V2I packets are largely outnumbered by BSM in terms of packet throughput. Because of this, the holistic measure that PER captures does not suit well for measuring service performance. We utilize PER to understand the impact of the service on the situational awareness within the network.   
%%%%%%%%%%%%%%%%%%%%%%%%%%%%%%%%%%%%%%%%%%%%%%%%%%%%%%%%%%%%%%%%%%

\section{Analysis \& Results}
\begin{figure}
\centerline{\includegraphics[width=0.5\textwidth,height=5cm]{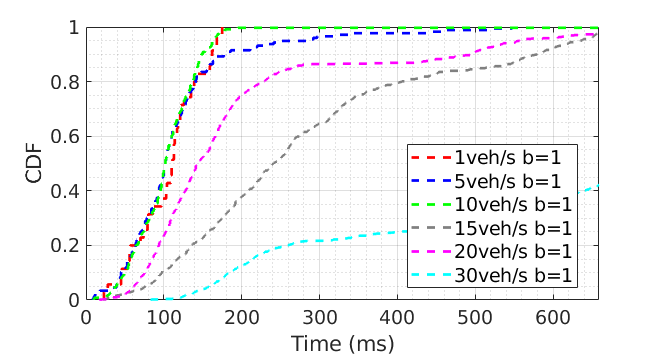}}
\caption{SCT for baseline (b=1) across traffic flow rates (1-30veh/s)}
\label{fig_ALL_b1}
\end{figure}

\begin{figure}[t]
\centerline{\includegraphics[width=0.5\textwidth,height=5cm]{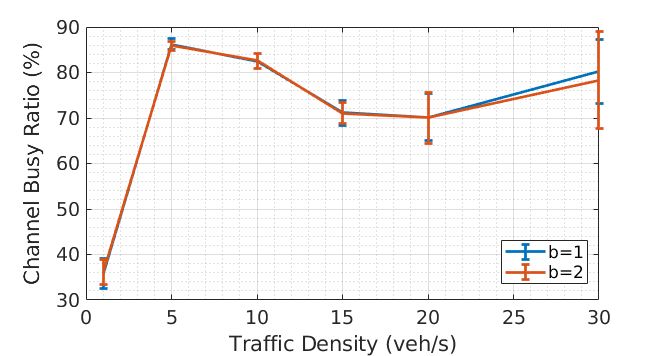}}
% \centerline[width=0.5\textwidth,height=6cm]{cbr-15_20_30.png}
\caption{CBR for batchsizes (b=1,2) across traffic flow rates (1-30veh/s)}
\label{fig_cbr}
\end{figure}

% We present two perspectives for analysing the system.
% \textbf{1. Service Utility Perspective} \\
% \textbf{2. Network Quality Perspective}  \\
In this section we explore the experiment findings to measure the efficiency of the protocol in providing service, its scalability across different traffic flow rate, and its impact on the network quality for safety packet usage. We consider the instantaneous ACK response configuration (i.e. $b=1$) settings as the baseline for performance comparison. In figure \ref{fig_ALL_b1}, we present the SCT for very low, low, and low-medium traffic flow with baseline settings. We observe that for very low and low-medium traffic, 99\% vehicles are able to complete the procedure within 200ms. For low-medium scenario, some vehicles required an additional attempt to complete the procedure, causing a long SCT tail. This can happen due to packet loss for either of the two packets in the first attempt (SUM or ACK). However, at low density traffic (5veh/s), 80\% vehicles enjoy similar SCT as the very low and low-medium scenario from baseline settings, while 20\% suffers with a longer SCT tail. This plunge can be explained once we observe Channel Busy Ratio (CBR) in these scenarios, presented in figure \ref{fig_cbr}. 

CBR indicates a relative measure of the channel status. At any subframe $t$, a VUE seeking a future transmission opportunity monitors the [$t, t-100$] ms window in its reception history, and computes CBR based on the following equation:
\begin{equation}
C B R(\%)=100 \times \frac{N_{R S S I}>\text { CBIThresh }}{N_{\text {total }}}
\end{equation}
% \[ CBR (\%) = 100 \times \frac{N_{RSSI>CBRThresh}}{N_{total}} \]
Here, $N_{RSSI>CBRThresh}\in \mathbb{N}$ is the number of resources within the specified window, that have a Reference Signal Received Power (RSRP) higher than a predefined CBR threshold. These resources are perceived as occupied by the receiver. A ratio of this quantity to the total number of resources available within the 100 subframes yields the relative measure of channel business. We emphasize on observing the CBR mainly for two reasons: 
\begin{itemize}
\item CBR works as the channel quality indicator at any given time of operation,
\item It is measured using locally available information at medium access control layer during onboard operations of a VUE, hence this can be closely tied to SCT.
\end{itemize}

\begin{figure}[t]
\centerline{\includegraphics[width=0.5\textwidth,height=5cm]{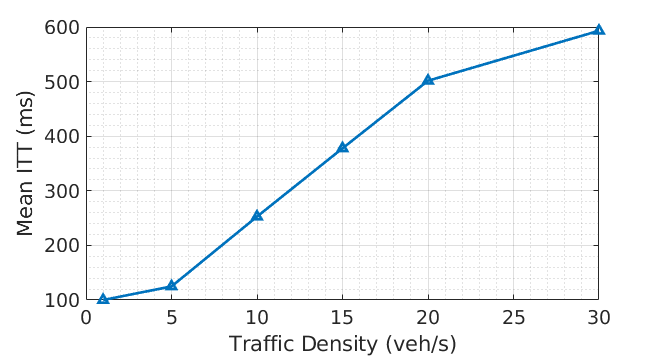}}
% \centerline[width=0.5\textwidth,height=6cm]{cbr-15_20_30.png}
\caption{Mean ITT for baseline (b=1) across traffic flow rates (1-30veh/s)}
\label{fig_itt}
\end{figure}

During on-board operation of a VUE, CBR is utilized in Rate Control (RC) algorithm to minimize congestion in distributed mode. We adopted the Rate Control (RC) algorithm specified in SAE J3161/1 \cite{saej3161} for our experiments. When traffic flow is high, RC allows a VUE to sense the channel status and adjust it's BSM periodicity accordingly. Channel status signifies the traffic density in the VUE's surrounding vicinity, which is effectively estimated from CBR. When high CBR is sensed, it decreases the BSM transmission rate, thereby increases the Inter-Transmitting Time (ITT). Mean ITT for the tested scenarios is presented in figure \ref{fig_itt}, which heavily correlates with the CBR at corresponding densities. At very low traffic (1veh/s), 35\% CBR (figure \ref{fig_cbr}) causes RC to remain dormant, resulting in VUEs transmitting BSMs with ~100ms ITT (figure \ref{fig_itt}). But a slight increase of traffic flow (5veh/s) vastly increases the CBR to ~90\%. This change is not reflected comparably at ITT, which barely increases to ~110ms. This low sensitivity of RC algorithm at low traffic flow appears from the algorithm originally being optimized for the older DSRC technology, which follows a different abstraction of physical layer bandwidth than C-V2X. The shortcomings of standard RC algorithm on C-V2X has been explored in \cite{toghi2019analysis}, which is highly suggested for readers interested in congestion control approaches in C-V2X. Further increase in density (10 veh/s) causes gradual increase in ITT. With higher ITT, larger volume of resources are available within a unit time-frame due to less frequent transmission by the VUEs. As a result, CBR comparably decreases and the now-accessible resources are utilized by the service application, making SCT at 10veh/s better than 5veh/s (figure \ref{fig_ALL_b1}). 

% (REFER CR Figure) CBR maintains resource access allowance across different application services to limit the channel occupancy allowed for each set of service  (REFER CR Figure)  (REFER CR Figure)  (REFER CR Figure)  (REFER CR Figure)  (REFER CR Figure)
At high CBR, SPS resource allocation procedure is more aggressive in resource shortlisting and selection. When a VUE transmitting in a high density traffic scenario increases its SPS RSRP threshold while shortlisting a resource for its own transmission, one implication of this occurrence is that the VUE can likely select a resource with low RSRP for its upcoming transmission. A resource with low RSRP is likely to be a resource that is currently being used by a vehicle situated far from the transmitter. However, when the transmitter utilizes that selected resource, the packet decodability is higher for the nearer transmitter's packet than the far one. This causes a receiver to decode the nearer packet and discard the far packet. As traffic density increases, free available resource grows more and more scarce, thus the chance of overlapping selection between a far transmitter (low RSRP) and a near transmitter (high RSRP) increases. Consequently, the effective range for a receiver VUE shortens as traffic density increases. While SPS is the de facto resource allocation procedure for C-V2X, such events benefit V2I packet receptions over BSM. V2I links for the service occurs between entities at small distance (RSU and VUE), whereas BSM links can be established between any two pair of entities within network coverage (function of density, road topology etc). When SPS aids reception of closer packet through the abovementioned events, V2I packets are the prime beneficiary who enjoys higher successful reception. This V2I-specific benefit and the reduction of range works counteractively in the low-medium to high traffic flow span, causing superior SCT at high traffic flow. At very high (30veh/s) traffic flow however, RC algorithm reaches its upper limit of ITT at 600ms, and the scope of more available resources due to range reduction diminishes, causing drastic impact on SCT.  

At the same time, the V2I packets themselves can pose detrimental effect on the V2I transmissions due to the following reasons:
\begin{itemize}
\item High traffic flow rate inadvertently generates high volume of SUM and ACK. Since all the SUM and ACK communications take place nearby RSU within a short distance, their perceived channel state have similar profile, resulting in similar shortlisted resources. These SUM transmissions are triggered in their individual first opportunity after crossing the trigger line, which makes it difficult to sense each other within their individual resource allocation procedure. This can increase the chance of collision between SUMs. Collision with ACK are equally likely as well since the RSU schedules ACK responses following similar resource allocation procedure. 
\item All C-V2X entities (VUE, RSU alike) suffer from half-duplex problem, which is even more challenging to tackle under distributed mode of operation (i.e. mode-4 sidelink operation). For V2I, half-duplex can mean UEs losing an ACK while transmitting SUM, for RSUs this can mean losing SUMs while transmitting a SAM or ACK. Missing either a SUM or an ACK will cause the failing VUE to re-transmit a SUM and keep on repeating until completion. At high density, these losses can initiate a spiraling increase in V2I transmissions.
\item RSUs have limited capability in transmitting multiple consequent ACK within short duration. Whether this poses an additional delay in service depends on the relative differences in traffic flow rate and service rate. If the traffic flow rate is higher than the RSU's service rate, the received SUMs form a request queue at the lower layers of RSU consisting of ACKs awaiting for transmission.
\end{itemize}

\begin{figure}[t]
\centerline{\includegraphics[width=0.5\textwidth,height=5cm]{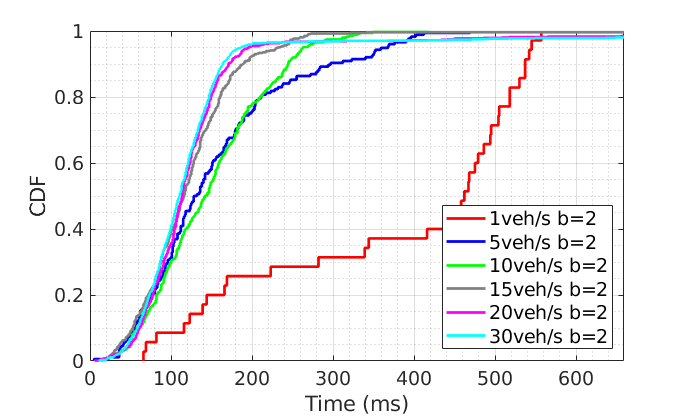}}
% \centerline[width=0.5\textwidth,height=6cm]{cbr-15_20_30.png}
\caption{SCT for (b=2) across traffic flow rates (1-30veh/s)}
\label{fig_ALL_b2}
\end{figure}

\begin{figure}[t]
\centerline{\includegraphics[width=0.5\textwidth,height=5.5cm]{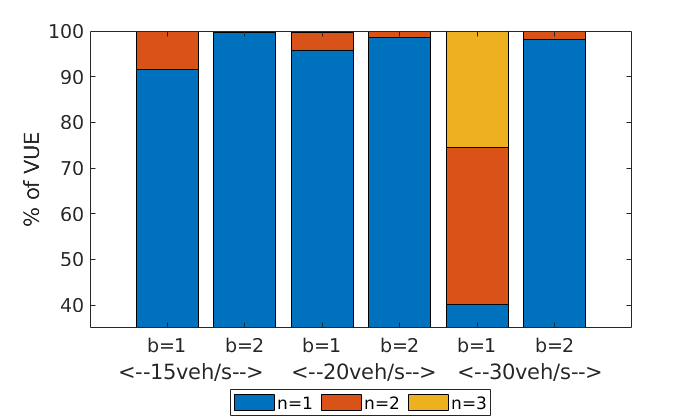}}
\caption{Percentage of VUEs with service completion at n-th try at different scenarios: b=(1,2) in medium to high traffic flow rates (15,20,30 veh/s)}
\label{fig_percent_completion_b1_b2}
\end{figure}

% talk about b=2
Under batched ACK settings, due to the nature of batched ACK, two additional factors come into play for a SUM-ACK cycle to be successful. 
\begin{itemize}
\item In each batch, the first member is subject to longer SCT than its latter members, as it transmitted earlier but has to wait until the last SUM is received by the RSU to fill up the last spot in that batch. To put it into comparison, only the last member in each batch enjoys a SCT comparable with its equivalent baseline scenario, while all the previous members in the same batch suffers with a slightly longer delay. The earlier the SUM is, the longer the SCT is compared to baseline for that particular VUE.  
\item Secondly, since each ACK contains $b$ number of responses, a loss of one ACK due to any of the reasons discussed above means $b$ number of VUE will repeat their SUM attempts. This means the channel can be congested with $b$ number of new messages. The actual number of increased message count could be higher since in a high traffic flow scenario, this can potentially worsen the spiraling effect and cause the loss of even more packets. 
\end{itemize}

While the baseline settings can provide the VUEs fast service at low traffic flow, it is certainly not true at high traffic flow. On the other hand, batched ACK can reduce the number of packets exchanged but it risks causing higher number of repeated attempts. We observed that $b=2$ settings can fuse the benefits from both notion. Figure \ref{fig_ALL_b2} shows that while scenarios in the very low to low-medium span suffers with additional delay in SCT, medium to very high scenario span thrives under the same settings. Figure \ref{fig_percent_completion_b1_b2} tracks the individual attempts by the same VUEs and their corresponding success in service completion. It shows that most VUE were able to complete the service with the second attempt, with a minor percentage requiring the third attempt in case of very high traffic.

\begin{figure}[t]
\centerline{\includegraphics[width=0.5\textwidth,height=7cm]{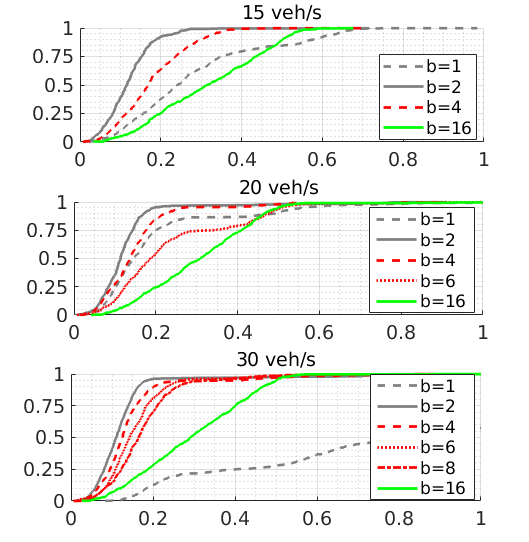}}
% \centerline[width=0.5\textwidth,height=6cm]{cbr-15_20_30.png}
\caption{SCT with different batchsizes for medium, high and very high traffic flow rate}
\label{fig_ttt_high}
\end{figure}

For the medium, high, and very high traffic flow scenarios, we investigated larger batchsizes seeking optimality (figure \ref{fig_ttt_high}). However, $b=2$ appears to remain the most optimal settings even in those scenarios. In larger batchsize, all batch members except the last one has relatively higher SCT than corresponding $b=2$ cases. The accumulation of these delays worsens the resulting SCT, causing a rightward shift. The $b=16$ settings produces a linearly increasing SCT line in all scenarios as this batchsize is much larger than the actual max number of incoming VUEs  within each ACK transmission interval at any of the traffic flow rates under test. 

\begin{figure}[t]
\centerline{\includegraphics[width=0.5\textwidth,height=7cm]{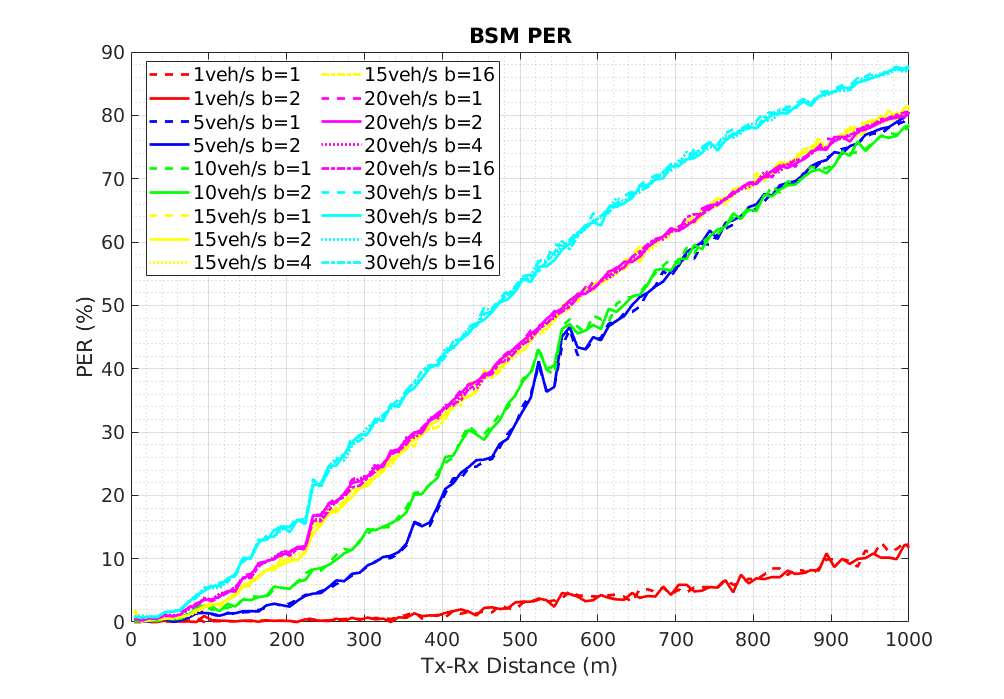}}
% \centerline[width=0.5\textwidth,height=6cm]{cbr-15_20_30.png}
\caption{Packet Error Rate (PER) under with different batchsize settings under very low to very high traffic flow rate (1-30 veh/s)}
\label{fig_per}
\end{figure}

We observed the impact of batchsize on the reception of BSMs as the advanced safety services are expected to leave basic safety communication unharmed. In figure \ref{fig_per}, PER shows minor influence resulting from different batchsize settings.

\section{Concluding Remarks and Future Outlook}
Infrastructure-based communication services are among the fundamental blocks for the next generation traffic system. To enable seamless relay between mobile traffic and stationary infrastructure entities, we explored multiple acknowledgement strategies that can be used in advanced safety and automated driving services. Resulting service performance and reception status show that the deployments should avoid static batching policy for acknowledgement packets, rather a scheme that adjusts in accordance with traffic condition can ensure quality of service. In addition, this paper indicates the issues regarding medium access for aperiodic packets and coexistence of multi-priority packets, that should be further investigated while C-V2X advances as the de facto VANET protocol.

\bibliographystyle{ieeetr}
\bibliography{references.bib}

\end{document}